\documentclass{PoS}

\title{NNLO real corrections to gluon scattering}

\ShortTitle{NNLO real corrections to gluon scattering}

\author{\speaker{Joao Pires}, E.W.N. Glover%
         \\
        Institute for Particle Physics Phenomenology\\
	Department of Physics\\
	University of Durham\\
	DH1 3LE\\
	UK\\
        E-mail: \email{joao.pires@durham.ac.uk}, 	\email{e.w.n.glover@durham.ac.uk}}

\abstract{In this talk we describe a procedure for isolating the infrared singularities present in gluonic scattering amplitudes at  next-to-next-to-leading order. We use the antenna subtraction framework which has been successfully applied to the calculation of NNLO corrections to the 3-jet cross section and related event shape distributions in electron-positron annihilation. Here we consider processes with coloured particles in the initial state, and in particular two-jet production at hadron colliders such as the Large Hadron Collider (LHC). We construct a subtraction term that describes
the single and double unresolved contributions from gluonic processes using
antenna functions with initial state partons and show numerically that the subtraction term correctly approximates the matrix elements in the various single and double unresolved configurations.}

\FullConference{RADCOR 2009 - 9th International Symposium on Radiative Corrections (Applications of Quantum Field Theory to Phenomenology) \\
                 October 25-30 2009\\
                 Ascona, Switzerland}

\begin{document}
\section{Introduction}
In hadronic collisions, the most basic form of the strong interaction at short distances is the scattering of a coloured parton off another coloured parton.  Experimentally, such scattering can be observed via the production of one or more jets of hadrons with with large transverse energy. In QCD, the scattering cross section has the perturbative  expansion,
\begin{equation}
{\rm d}\sigma =\sum_{i,j} \int  \left[
{\rm d}\hat\sigma_{ij}^{LO}
+\left(\frac{\alpha_s}{2\pi}\right){\rm d}\hat\sigma_{ij}^{NLO}
+\left(\frac{\alpha_s}{2\pi}\right)^2{\rm d}\hat\sigma_{ij}^{NNLO}
+{\cal O}(\alpha_s^3)
\right]f_i(x_1) f_j(x_2) dx_1 dx_2 \nonumber
\end{equation}
where the sum runs over the possible parton types $i$ and $j$.   The single-jet inclusive and di-jet cross sections have been studied at next-to-leading order (NLO) ~\cite{NLOjet} and successfully compared with data from the TEVATRON. 

The theoretical prediction may be improved by including the
next-to-next-to-leading order (NNLO) perturbative predictions.  This has the effect of (a) reducing the renormalisation scale dependence and (b) improving the matching of the parton level theoretical jet algorithm with the hadron level
experimental jet algorithm because the jet structure can be modeled by the
presence of a third parton.  The resulting theoretical uncertainty
at NNLO is estimated to be at the few per-cent level~\cite{Glover:2002gz}. 

In this talk, we will focus only on the NNLO contribution involving gluons and will drop the parton labels.
At NNLO, there are three distinct contributions due to double real radiation radiation ${\rm{d}}\sigma_{NNLO}^R$, mixed real-virtual radiation ${\rm{d}}\sigma_{NNLO}^{V,1}$ and double virtual radiation ${\rm{d}}\sigma_{NNLO}^{V,2}$, that are given by
\begin{eqnarray}
{\rm d}\hat\sigma_{NNLO}&=&\int_{{\rm{d}}\Phi_{m+2}} {\rm{d}}\hat\sigma_{NNLO}^R 
+\int_{{\rm{d}}\Phi_{m+1}} {\rm{d}}\hat\sigma_{NNLO}^{V,1} 
+\int_{{\rm{d}}\Phi_m}{\rm{d}}\hat\sigma_{NNLO}^{V,2}
\end{eqnarray}
where the integration is over the appropriate $N$-particle final state subject to the constraint that precisely $m$-jets are observed,
\begin{equation}
\int_{{\rm{d}}\Phi_{N}} = \int {\rm{d}}\Phi_{N} J_m^{(N)}.
\end{equation}
As usual the individual contributions in the $(m+2)$, $(m+3)$ and $(m+4)$-parton channels are all separately infrared divergent although, after renormalisation and factorisation, their sum is finite. 
For processes with two partons in the initial state, the parton level cross sections are related to the interference of $M$-particle $i$-loop and $j$-loop amplitudes $[\langle{\cal M}^{(i)}|{\cal M}^{(j)}\rangle]_M$ by
\begin{eqnarray}
{\rm{d}}\hat\sigma_{NNLO}^R &\sim& \left[\langle{\cal M}^{(0)}|{\cal M}^{(0)}\rangle\right]_{m+4}\nonumber\\
{\rm{d}}\hat\sigma_{NNLO}^{V,1} &\sim& \left[\langle{\cal M}^{(0)}|{\cal M}^{(1)}\rangle+\langle{\cal M}^{(1)}|{\cal M}^{(0)}\rangle\right]_{m+3}\nonumber\\
{\rm{d}}\hat\sigma_{NNLO}^{V,2} &\sim& \left[\langle{\cal M}^{(1)}|{\cal M}^{(1)}\rangle+\langle{\cal M}^{(0)}|{\cal M}^{(2)}\rangle+\langle{\cal M}^{(2)}|{\cal M}^{(0)}\rangle\right]_{m+2}
\end{eqnarray}
In this talk, we specialise to the gluonic contributions to dijet production.
Explicit expressions for the interference of the four-gluon tree-level and two-loop amplitudes is available in Refs.~\cite{gggg}, while the self interference of the four-gluon one-loop amplitude is given in \cite{Glover:2001rd}. The one-loop helicity amplitudes for the five gluon amplitude are given in \cite{Bern:1993mq}. This contribution contains explicit infrared divergences coming from integrating over the loop momenta and implicit poles in the regions of the phase space where one of the final state partons becomes unresolved. This corresponds to the soft and collinear regions of the one-loop amplitude that were analyzed in \cite{Bern:1998sc}.
The double real  six-gluon matrix elements were derived in \cite{gggggg}. Here the singularities occur in the phase space regions corresponding to two gluons becoming simultaneously soft and/or collinear. The ``double'' unresolved behaviour is universal and was discussed in \cite{Campbell:1997hg, Catani:1998nv}.

\section{The antenna subtraction formalism}

There have been several approaches to build a general subtraction scheme for the double real contribution at NNLO \cite{Weinzierl:2003fx,Frixione:2004is,Somogyi:2005xz}. We will follow the antenna subtraction method which was derived in \cite{GehrmannDeRidder:2005cm} for NNLO processes involving only (massless) final state partons. This formalism is being extended to include processes with either one or two initial state partons   ~\cite{Daleo:2006xa,Daleo:2009yj,Boughezal:2010ty}.


To render the contributions with different final states separately finite, 
the general structure of the subtraction terms at NNLO is
\begin{eqnarray}
{\rm d}\hat\sigma_{NNLO}&=&\int_{{\rm{d}}\Phi_{m+2}}\left({\rm{d}}\hat\sigma_{NNLO}^R-{\rm{d}}\hat\sigma_{NNLO}^S\right)
+\int_{{\rm{d}}\Phi_{m+2}}{\rm{d}}\hat\sigma_{NNLO}^S\nonumber\\
&+&\int_{{\rm{d}}\Phi_{m+1}}\left({\rm{d}}\hat\sigma_{NNLO}^{V,1}-{\rm{d}}\hat\sigma_{NNLO}^{V S,1}\right)
+\int_{{\rm{d}}\Phi_{m+1}}{\rm{d}}\hat\sigma_{NNLO}^{V S,1}\nonumber\\
&+&\int_{{\rm{d}}\Phi_m}{\rm{d}}\hat\sigma_{NNLO}^{V,2},
\end{eqnarray}
where ${\rm{d}}\hat\sigma_{NNLO}^S$ (${\rm{d}}\hat\sigma_{NNLO}^{V S,1}$) is the subtraction term for the double radiation (real-virtual) contributions respectively.

In this talk, we concentrate on the double unresolved subtraction term ${\rm{d}}\hat\sigma_{NNLO}^S$ relevant for the six-gluon contribution two-jet production in hadronic collisions.  It is made up of several different contributions, that depend on how the unresolved partons are connected in colour space, 
\begin{eqnarray}
{\rm d}\hat\sigma_{NNLO}^S={\rm d}\hat\sigma_{NNLO}^{S,a}+{\rm d}\hat\sigma_{NNLO}^{S,b}+{\rm d}\hat\sigma_{NNLO}^{S,c}+{\rm d}\hat\sigma_{NNLO}^{S,d}+{\rm d}\hat\sigma_{NNLO}^{A}\nonumber
\end{eqnarray}
\begin{itemize}
\item[(a)] One unresolved parton but the experimental observable selects only
$m$ jets from the $(m+1)$ partons.   
\item[(b)] Two colour-connected unresolved partons (colour-connected).
\item[(c)] Two unresolved partons that are not colour connected but share a common
radiator (almost colour-unconnected).
\item[(d)] Two unresolved partons that are well separated from each other 
in the colour 
chain (colour-unconnected).
\item[(A)] Correction for the over subtraction of large angle soft radiation.
\end{itemize}

Contribution (a) is precisely the same subtraction term as used for the NLO $(m+1)$-jet rate, and is the product of a three-parton antenna and reduced 
$(m+1)$-particle matrix elements.  Contributions (b)--(d) are derived from the
product
of double unresolved factors and 
reduced $(m+2)$-parton matrix elements. Subtraction terms for these 
configurations can be constructed using either 
single four-parton antenna functions~\cite{GehrmannDeRidder:2005hi} or products of three-parton antenna 
functions. For gluonic processes, one encounters the four-gluon antenna $F_{4}^{0}$ for the first time. Finally, the  large-angle soft subtraction terms (A) contains 
soft antenna functions which precisely cancel the remnant soft behaviour  associated with the
antenna phase space mappings for the final-final \cite{Weinzierl:2008iv}, initial-final and initial-initial configurations. Explicit formulae for the various contributions to ${\rm d}\hat\sigma_{NNLO}^S$ are available in \cite{Glover:2010}

Note that the subtraction terms are also needed in integrated form. When both radiators are in the final state, as needed for electron-positron annihilation, the integrated antennae are given in ref.~\cite{GehrmannDeRidder:2005cm}.   For processes with one hard radiator in the initial state, the integrals are known \cite{Daleo:2009yj} while the work is still in progress for processes with two hadronic initial radiators \cite{Boughezal:2010ty}.

\section{Numerical results}

To numerically test that the subtraction term correctly reproduces the
same infrared behaviour as the matrix element, we generate a series of phase space points that approach a given double or single unresolved limit. For each generated point we compute the ratio,
\begin{equation}
R=\frac{{\rm d}\hat\sigma_{NNLO}^R}{{\rm d}\hat\sigma_{NNLO}^S}
\end{equation}
which should approach unity as we get closer to any singularity.

\begin{figure}[t]
\begin{minipage}[b]{0.3\linewidth}
\centering
\includegraphics[width=4cm]{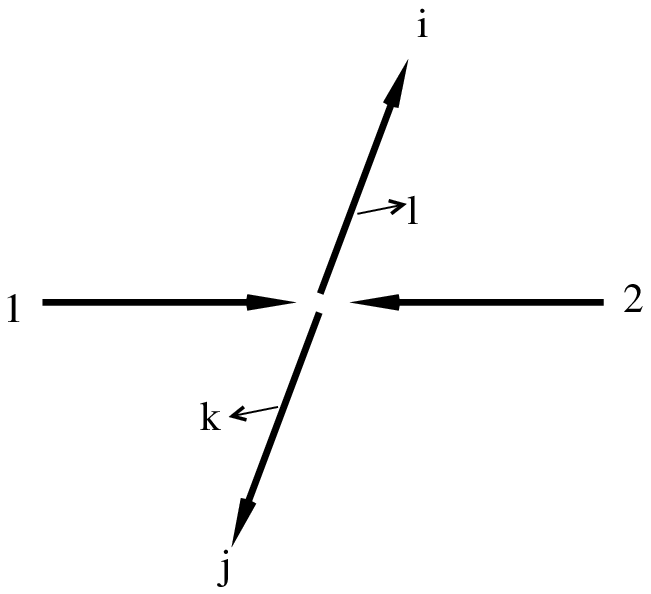}\\
\vspace{0.2cm}
(a)
\end{minipage}
\hspace{0.5cm}
\begin{minipage}[b]{0.7\linewidth}
\centering
\includegraphics[width=5cm,angle=270]{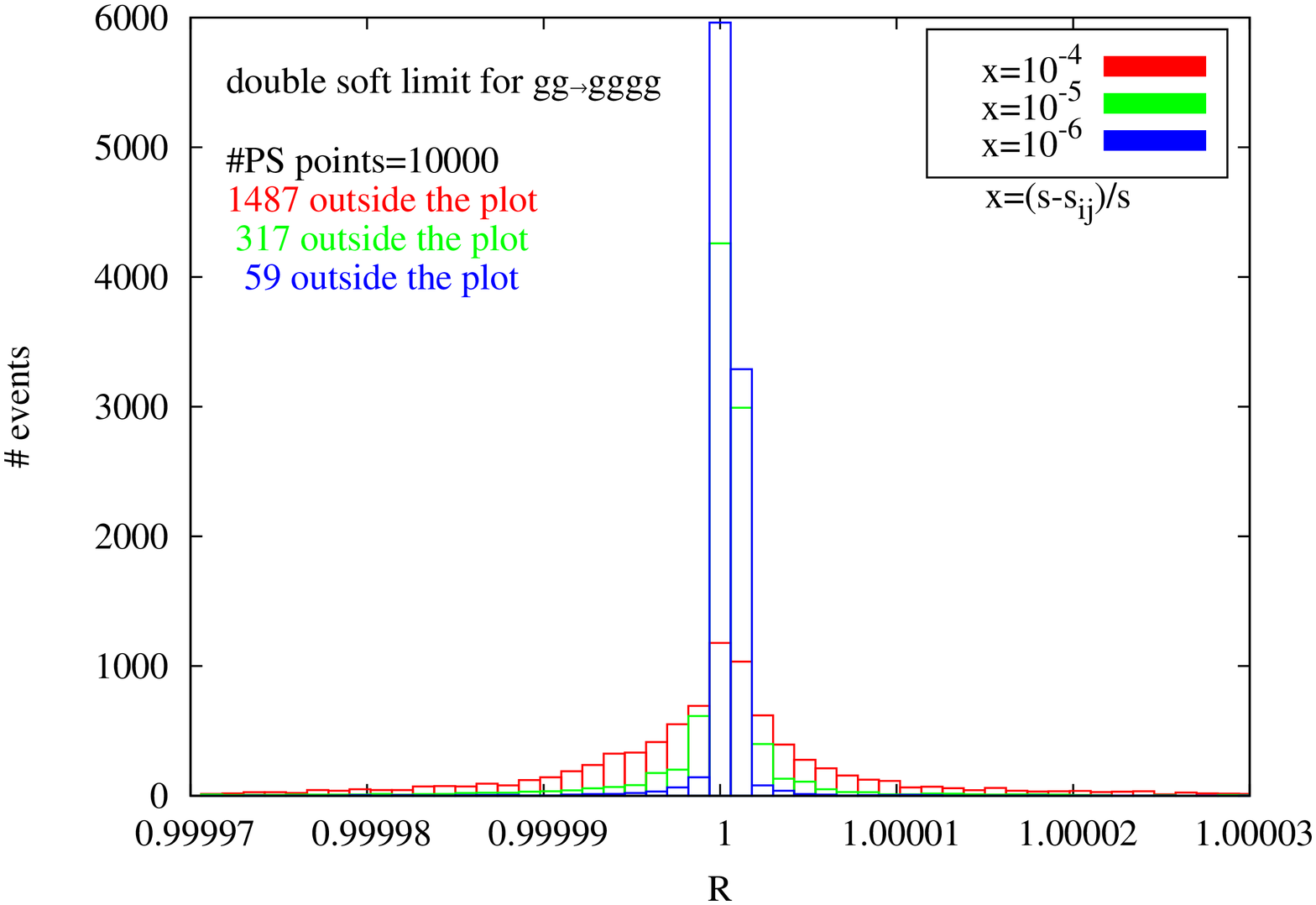}\\
\vspace{0.2cm}
(b)
\end{minipage}
\caption[Double soft limit distributions]{(a) Example configuration of a double soft event with $s_{ij}\approx s_{12}=s$.
(b) Distribution of $R$ for 10000 double soft phase space points.}
\label{fig:dsoft}
\end{figure}

As an example, we consider the double soft limit.
A double soft configuration can be obtained by generating a four particle final state where one of the invariant masses $s_{ij}$ of two final
state particles takes nearly the full energy of the event $s$ as illustrated in figure \ref{fig:dsoft} (a).

In figure \ref{fig:dsoft}(b) we generated 10000 random double soft phase space points and show the
distribution of the ratio between the matrix element and the subtraction term. We show three different values of $x=(s-s_{ij})/s$ [$x=10^{-4}$ (red), $x=10^{-5}$ (green), $x=10^{-6}$ (blue)] and we can see that for smaller values of $x$ the distribution peaks more sharply around unity. For $x=10^{-6}$ we obtained an average of $R=0.9999994$ and a standard deviation of $\sigma=4.02\times10^{-5}$. Also in the plot we give the number of outlier points that lie outside the range of the histogram. As expected this number decreases as we approach the singular region. 

\begin{figure}[t]
\begin{minipage}[b]{0.3\linewidth}
\centering
\includegraphics[width=4cm]{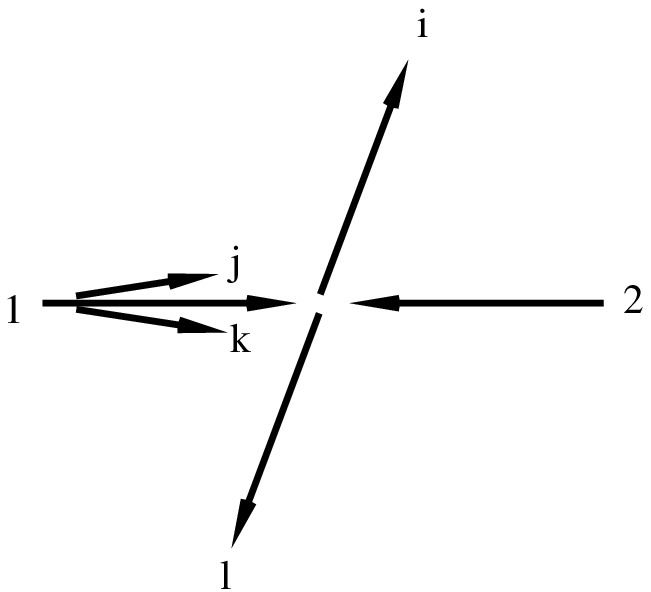}\\
\vspace{0.2cm}
(a)
\end{minipage}
\hspace{0.5cm}
\begin{minipage}[b]{0.7\linewidth}
\centering
\includegraphics[width=5cm,angle=270]{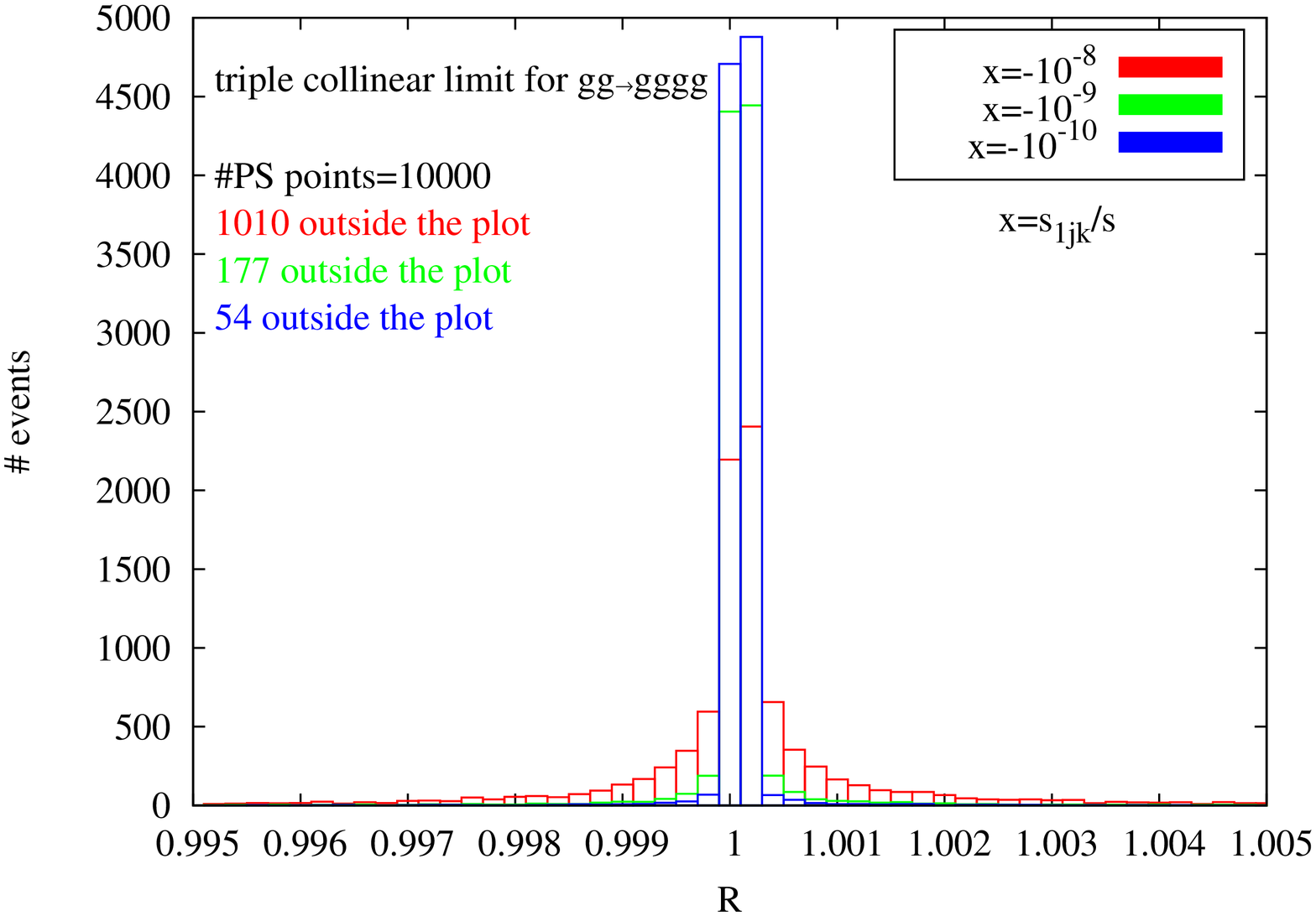}\\
\vspace{0.2cm}
(b)
\end{minipage}
\caption[Triple collinear limit initial state singularity]{(a) Example configuration of a triple collinear event with $s_{1jk}\to 0$.
(b) Distribution of $R$ for 10000 triple collinear phase space points.}
\label{fig:tcolliff}
\end{figure}

As a second example, we perform a similar analysis for the triple collinear limit with three hard particles sharing a collinear direction as shown in figure
\ref{fig:tcolliff} (a).  In this case the variable that controls the approach to the triple collinear region is $x=s_{1jk}/s$.
We show results for $x=-10^{-7}$ (red), $x=-10^{-8}$ (green), $x=-10^{-9}$ (blue).  For 10000 phase space points with $x=-10^{-9}$, 
we obtained an average value of $R=0.99954$ and a standard deviation of $\sigma=0.04$. As before, the number of outliers systematically decreases as we approach the triple collinear limit.

Similar behaviour is obtained for all of the remaining double unresolved limits.

\section{Conclusions}

In this contribution, we have discussed the application of the antenna
subtraction formalism to construct the subtraction term relevant for the gluonic
double real radiation contribution to dijet production. The subtraction term is
constructed using four-parton and three-parton antennae.  We showed that the
subtraction term correctly describes the double unresolved limits of the $gg \to
gggg$ process. The construction of similar subtraction terms for processes
involving quarks should in principle be straightforward.   Together with the integrated
forms of the antenna functions (see Refs.~\cite{Daleo:2009yj} for
the initial-final and  Ref.~\cite{Boughezal:2010ty} for the initial-initial
configurations), these double real subtraction terms will provide a major step towards the
NNLO evaluation of the dijet  observables at hadron colliders. 

\section{Acknowledgements}

This research was supported in part by the UK Science and Technology Facilities
Council and by the European Commission's Marie-Curie Research Training Network
under contract MRTN-CT-2006-035505 `Tools and Precision Calculations for Physics
Discoveries at Colliders'. EWNG gratefully acknowledges the support of the
Wolfson Foundation and the Royal Society. JP gratefully acknowledges the award
of a Funda\c{c}\~{a}o para a Ci\^encia e Tecnologia (FCT - Portugal) PhD
studentship.


\end{document}